\documentstyle[12pt,epsf]{article}

\newfont{\lfont}{line10}

\newcommand{\BE}{\begin{equation}}
\newcommand{\EE}{\end{equation}}
\newcommand{\BEA}{\begin{eqnarray}}
\newcommand{\EEA}{\end{eqnarray}}

\newcommand{\JHEP}[3]{JHEP~{\bf #1}{(#2)}{#3}}

\newcommand{\JMP}[3]{J. Math. Phys. {\bf #1}{(#2)}{#3}}
\newcommand{\NP}[3]{Nucl. Phys. {\bf #1}{(#2)}{#3}}
\newcommand{\PL}[3]{Phys. Lett. {\bf #1}{(#2)}{#3}}
\newcommand{\PR}[3]{Phys. Rev. {\bf #1}{(#2)}{#3}}

\newcommand{\PTP}[3]{Prog. Theo. Phys. {\bf #1}{(#2)}{#3}}

\def\12{\frac{1}{2}}
\def\bea{\begin{eqnarray}}
\def\eea{\end{eqnarray}}
\def\ba{\begin{array}}
\def\ea{\end{array}}

\def\one-loop{\mbox{\scriptsize one-loop}}

\def\G{\Gamma}

\catcode`\@=11

\@addtoreset{equation}{section}
\def\theequation{\arabic{section}.\arabic{equation}}

\def\@normalsize{\@setsize\normalsize{15pt}\xiipt\@xiipt
\abovedisplayskip 14pt plus3pt minus3pt%
\belowdisplayskip \abovedisplayskip
\abovedisplayshortskip  \z@ plus3pt%
\belowdisplayshortskip  7pt plus3.5pt minus0pt}

\def\small{\@setsize\small{13.6pt}\xipt\@xipt
\abovedisplayskip 13pt plus3pt minus3pt%
\belowdisplayskip \abovedisplayskip
\abovedisplayshortskip  \z@ plus3pt%
\belowdisplayshortskip  7pt plus3.5pt minus0pt
\def\@listi{\parsep 4.5pt plus 2pt minus 1pt
            \itemsep \parsep
            \topsep 9pt plus 3pt minus 3pt}}

\def\underline#1{\relax\ifmmode\@@underline#1\else
        $\@@underline{\hbox{#1}}$\relax\fi}
\@twosidetrue

\relax

\catcode`@=12

\catcode`\@=11

\def\section{\@startsection{section}{1}{\z@}{3.5ex plus 1ex minus
   .2ex}{2.3ex plus .2ex}{\large\bf}}

\def\thesection{\Roman{section}.}

\def\appendix{\setcounter{section}{0}
        \def\thesection{Appendix }
       \def\theequation{\Alph{section}.\arabic{equation}}}

\def\ps@headings{\def\@oddfoot{}\def\@evenfoot{}
\def\@oddhead{\hbox{}\hfill
        \makebox[.5\textwidth]{\raggedright\ignorespaces --\thepage{}--
        \hfill {}}}
\def\@oddhead{\hbox{}\hfill --\thepage{}-- \hfill
        {}}
\def\@evenhead{\@oddhead}
\def\subsectionmark##1{\markboth{##1}{}}
}

\ps@headings

\catcode`\@=12

\relax

\def\figcap{\section*{Figure Captions\markboth
        {FIGURECAPTIONS}{FIGURECAPTIONS}}\list
        {Fig. \arabic{enumi}:\hfill}{\settowidth\labelwidth{Fig. 999:}
        \leftmargin\labelwidth
        \advance\leftmargin\labelsep\usecounter{enumi}}}
 \relax
\def\tablecap{\section*{Table Captions\markboth
        {TABLECAPTIONS}{TABLECAPTIONS}}\list
        {Table \arabic{enumi}:\hfill}{\settowidth\labelwidth{Table 999:}
        \leftmargin\labelwidth
        \advance\leftmargin\labelsep\usecounter{enumi}}}
 \relax
\def\reflist{\section*{References\markboth
        {REFLIST}{REFLIST}}\list
        {[\arabic{enumi}]\hfill}{\settowidth\labelwidth{[999]}
        \leftmargin\labelwidth
        \advance\leftmargin\labelsep\usecounter{enumi}}}
 \relax

\catcode`\@=11

\def\ps@headings{\def\@oddfoot{}\def\@evenfoot{}
\def\@oddhead{\hbox{}\hfill
        \makebox[.5\textwidth]{\raggedright\ignorespaces --\thepage{}--
        \hfill {}}}
\def\@evenhead{\@oddhead}
\def\subsectionmark##1{\markboth{##1}{}}
}

\ps@headings

\relax

\newskip\humongous \humongous=0pt plus 1000pt minus 1000pt

\newif\ifdtup

\def\beq{\begin{equation}}
\def\eeq{\end{equation}}

\def\beqn{\begin{eqnarray}}
\def\eeqn{\end{eqnarray}}
\relax

\def\G2{{\; \rm GeV/}c2}
\def\G{\; \rm GeV}

\def\dotx{\dotx{\dot\overline{x}}}

\relax

\begin{document}
%
%
\begin{titlepage}

\renewcommand{\thefootnote}{\fnsymbol{footnote}}

\begin{flushright}
      \normalsize
  May, 2002 \\
  OCU-PHYS 186\\
  hep-th/0205163  \\

\end{flushright}

%
\begin{center}
  {\large\bf Off-shell Crosscap State and Orientifold Planes\\
  with Background Dilatons}
\footnote{This work is supported in part
 by the Grant-in-Aid  for Scientific Research
(14540264) from the Ministry of Education,
Science and Culture, Japan.}
\end{center}

\vfill

\begin{center}
    {%
Hiroshi~Itoyama\footnote{e-mail: itoyama@sci.osaka-cu.ac.jp}
\quad and \quad
Shin~Nakamura\footnote{e-mail: nakashin@postman.riken.go.jp}
}\\
\end{center}

\vfill

\begin{center}
      ${}^{\dag}$\it  Department of Mathematics and Physics,
        Graduate School of Science\\
        Osaka City University\\
        3-3-138, Sugimoto, Sumiyoshi-ku, Osaka 558-8585, Japan  \\

      ${}^{\ddag}$\it Theoretical Physics Laboratory,
      RIKEN (The Institute of Physical and Chemical Research)\\
       2-1 Hirosawa, Wako, Saitama 351-0198, Japan

\end{center}

\vfill


\begin{abstract}
We show that a non-trivial dilaton condensation alters
the dimensions of orientifold planes.
An off-shell crosscap state which naturally interpolates between 
the usual on-shell crosscap states and their T-duals plays an
important role in the analysis.
We present an explicit representation of the off-shell crosscap
state on an $RP^{2}$ worldsheet in the gauge in which the worldsheet
curvature in the bulk of the fundamental region of the $RP^{2}$ vanishes.
We show that the non-trivial dilaton condensation reproduces the
correct descent relation among orientifold plane tensions.

\end{abstract}

\vfill

\setcounter{footnote}{0}
\renewcommand{\thefootnote}{\arabic{footnote}}

\end{titlepage}
\section{Introduction}

Orientifold planes (O-planes) as well as D-branes are important
objects to reveal non-perturbative effects in unoriented string theory.
Recent studies on open-string tachyon condensation have shown
that background fields can control the dimensions of 
D-branes.\footnote{
See, for example, Refs. \cite{SFT-review} for recent reviews 
on the related topics.}
The relationship between the dimensions of the D-branes and the
configuration of the background open-string fields is easily
understood from the viewpoint of the worldsheet; the background 
fields on the boundaries can alter the boundary conditions 
on the worldsheet.
On the other hand, the relationship between the properties of O-planes 
and background fields has not been understood well as yet. 

In the present work, we investigate the connection between 
the dimensions of O-planes and the configuration of the 
background dilaton field, in unoriented bosonic string theory.
An O-plane is represented as a crosscap on an $RP^2$ worldsheet
while a D-brane is represented as a boundary on a disc.
Therefore we consider an $RP^2$ worldsheet in the presence of the
background dilatons.
The reason we consider dilatons is based on the following property.
Dilatons couple to the worldsheet curvature and their 
contribution can be put on any part of the worldsheet
in general.
However, we show that the contribution of the background
dilatons localizes on the crosscap if we 
choose the gauge in which the worldsheet curvature vanishes 
in the bulk of the fundamental region of the $RP^2$.
This choice of gauge is nice; the bulk
part of the $RP^2$ becomes free and all the interactions
from the background dilatons appear only on the crosscap.
Therefore the effects of the background dilatons can be
translated into the modification of the crosscap conditions.

We introduce a new crosscap state which we refer to as an 
{\em off-shell crosscap state} in order to analyze 
the properties of the $RP^2$ worldsheet.
Of course, O-planes couple to only closed strings and we do not
have open-string modes on the $RP^2$ worldsheet.
Useful tools to analyze the properties of worldsheets in terms 
of closed-string modes are boundary states and crosscap states
\cite{Horava,Callan-etal-1,Callan-etal-2,Callan-etal-3}.
Usually, boundary states and crosscap states belong to the closed-string 
sector which preserves conformal invariance on the worldsheet.
Extensions of these boundary states which do not in general 
maintain conformal invariance have been proposed recently
\cite{Fujii-Itoyama-2,Lee}. 
(See also \cite{Akh-Laid-Seme,Laid-Seme}.)
We call them off-shell boundary states in the present article.
The off-shell boundary state interpolates between the usual 
on-shell boundary states and their T-duals.
This state is defined on a disc worldsheet with 
quadratic boundary interactions.
The dimension of the corresponding D-brane is controlled by the coupling
constants of these interactions. 
Boundary string field theory (BSFT) 
\cite{wittenbsft-1,wittenbsft-2} 
states that the coupling constants also parameterize 
the configuration of open-string tachyons
\cite{harvey-kutasov-Martinec,gerasimov-shatashvili,kutasov-marino-moore,0010108,0012198,0012210}.
Calculating the disc partition function by using these states enables
us to obtain the descent relation among the D-brane tensions 
if we take appropriate on-shell limits of the partition functions.

An attempt to apply the foregoing idea to crosscap states is given 
in Ref. \cite{Itoyama-nakamura}, in which
the definition of the off-shell crosscap state which interpolates the 
usual on-shell crosscap states and their T-duals has been proposed.
We present an explicit representation of the off-shell crosscap state
in the present work.
We define the off-shell crosscap state on an $RP^{2}$ worldsheet 
in the presence of quadratic interactions on the crosscap.
The dimension of the corresponding O-plane is controlled by the coupling
constants of these interactions.
The physical meaning of the quadratic interactions on the crosscap is
the background dilaton field of quadratic configuration. 
The behaviour of the off-shell crosscap state shows that the background
dilatons control the dimension of the O-plane.

The off-shell crosscap state is a useful tool to obtain
correlation functions on the $RP^2$ worldsheet in the presence
of the quadratic dilatons on the crosscap.
The exact partition function of the $RP^2$
worldsheet which is considered to be proportional to the O-plane
tension can be calculated exactly 
by using the two-point function.
We show that the condensation of the dilatons of quadratic profile 
reproduces the correct descent relation among O-plane tensions
by taking appropriate on-shell limits of the partition function. 
Note that we have not attempted to find the dynamical origin of the
dilaton condensation.

The paper is organized as follows.
In Sec. 2, we construct the $RP^2$ worldsheet on a complex plane and
we show that the contribution of the background dilatons survives 
on the crosscap alone if we choose the gauge in which the worldsheet
curvature vanishes in the bulk of the fundamental region of the $RP^2$.
In Sec. 3 we define the off-shell crosscap state and we obtain its
explicit representation.
We find that the behaviour of this state signifies that a non-trivial
dilaton condensation alters the dimensions of O-planes.
We calculate the partition function of the $RP^{2}$ by using the
off-shell crosscap state.
In Sec. 4, we verify that the non-trivial
dilaton condensation reproduces the correct descent relation of
O-plane tensions, by taking appropriate limits of the partition
function of the $RP^{2}$ worldsheet.
In the final section, we summarize the results of this study.
We make comments on some relationship between the $RP^{2}$ worldsheet and
the supersymmetric disc worldsheet which appears in the analysis
of $D\bar{D}$ systems in Appendix.

\section{$RP^{2}$ worldsheet with background dilatons}

Let us consider an $RP^{2}$ worldsheet with background dilaton field.
We show, in this section, that the contribution of 
the background dilatons localizes on the crosscap 
if we choose the gauge in which the worldsheet curvature 
in the bulk of the fundamental region of the $RP^2$ vanishes.

$RP^{2}$ is a non-orientable Riemann surface of Euler number one 
with no hole, no boundary and one crosscap.
We construct the $RP^{2}$ worldsheet on a complex $z$-plane 
by using an involution where we identify $z$ and $-\frac{1}{\bar{z}}$ 
on the complex plane.
We choose the fundamental region $\Sigma$ to be
$\{z=r e^{i\sigma} |0\le r < 1,0\le \sigma <2\pi\}\cup
\{z=r e^{i\sigma} |r= 1,0\le \sigma <\pi\}$.
The crosscap ${\cal C}$, the non-trivial closed loop of the $RP^{2}$
worldsheet, is represented as half of unit circle
$\{z=r e^{i\sigma} |r= 1,0\le \sigma <\pi\}$
in this case.

To begin with, we set the metric inside the unit circle
($|z| \le 1$) on the complex plane as
\begin{eqnarray}
h_{zz}=h_{\bar{z}\bar{z}}&=&0,
\nonumber \\
h_{\bar{z}z}=h_{z\bar{z}}&=&\frac{1}{2}.
\end{eqnarray}
The metric outside the unit circle ($|z'| \ge 1$) is obtained
by the involution $z'=-\frac{1}{\bar{z}}$, $\bar{z}'=-\frac{1}{z}$
as
\begin{eqnarray}
h_{z'\bar{z}'}
&=&
\frac{\partial \bar{z}}{\partial z'} 
\frac{\partial z}{\partial \bar{z}'}
h_{\bar{z}z}
\nonumber \\
&=&
\frac{1}{(z'\bar{z}')^{2}} h_{\bar{z}z}
=\frac{1}{r^{4}} h_{\bar{z}z},
\end{eqnarray}
where $r^{2}=z'\bar{z}'$ for $r \ge 1$.
Therefore the metric on the entire complex plane can be written as
\begin{eqnarray}
h_{zz}=h_{\bar{z}\bar{z}}&=&0,
\nonumber \\
h_{\bar{z}z}=h_{z\bar{z}}&=&\frac{1}{2} e^{\rho},
\end{eqnarray}
where
\begin{eqnarray}
\rho(r)=(-4\ln r) \: \theta(r-1).
\end{eqnarray}
Next, we rewrite the metric in the polar coordinate $(r,\sigma)$ as
\begin{eqnarray}
g_{ab}=\hat{g}_{ab}e^{\rho},
\end{eqnarray}
where 
$\hat{g}_{rr}=1,\hat{g}_{\sigma \sigma}=r^{2},
\hat{g}_{\sigma r}=\hat{g}_{r \sigma}=0$.
The worldsheet curvature $R$ is then given by
\begin{eqnarray}
\sqrt{g}R
&=&
-\sqrt{\hat{g}}\hat{\nabla}^{2}\rho(r)
=
-r
\left(
{\partial_{r}}^{2}+\frac{1}{r}\partial_{r}
\right) \rho(r)
\nonumber \\
&=&
4\left\{
\delta'(r-1) r\ln r +(2+\ln r)\delta(r-1)
\right\}.
\end{eqnarray}
Let us calculate the integral
$\int_{\Sigma(\epsilon)} dr d\sigma \sqrt{g}R \Phi(r,\sigma)$.
We define the integral region $\Sigma(\epsilon)$ as
$\{z=r e^{i\sigma} |0\le r \le 1+\epsilon,0\le \sigma <\pi\}\cup
\{z=r e^{i\sigma} |0\le r \le 1-\epsilon,\pi\le \sigma <2\pi\}$,
where $\epsilon$ is a positive real number.
$\Sigma(\epsilon)$ becomes the fundamental region $\Sigma$ of the
$RP^2$ after we take the limit $\epsilon \rightarrow 0$.
We obtain
\begin{eqnarray}
\int_{\Sigma(\epsilon)} dr d\sigma \sqrt{g}R \Phi(r,\sigma)
&=&
\int_{0}^{\pi} d\sigma \int_{0}^{1+\epsilon} dr
\sqrt{g}R \Phi(r,\sigma)
+
\int_{\pi}^{2\pi} d\sigma \int_{0}^{1-\epsilon} dr
\sqrt{g}R \Phi(r,\sigma)
\nonumber \\
&=&4
\int_{0}^{\pi} d\sigma \int_{0}^{1+\epsilon} dr
\left\{
\delta'(r-1) r\ln r + (2+\ln r)\delta(r-1)
\right\}\Phi(r,\sigma)
\nonumber \\
&=&
4
\int_{0}^{\pi} d\sigma
\left\{
-\frac{d}{dr}\{r\ln r \:\:\Phi(r,\sigma)\}|_{r=1}
 + 2\Phi(1,\sigma)
\right\}
\nonumber \\
&=&
4
\int_{0}^{\pi} d\sigma
\Phi(1,\sigma)
,
\end{eqnarray}
which yields
\begin{eqnarray}
\frac{1}{4\pi}
\int_{\Sigma} dr d\sigma \sqrt{g}R \Phi(r,\sigma)
=
\frac{1}{\pi} \int_{0}^{\pi} d\sigma
\Phi(1,\sigma).
\label{dilaton-RP2}
\end{eqnarray}
Note that (\ref{dilaton-RP2}) gives the correct Euler number of 
the $RP^2$ (which is one) if we set $\Phi=1$.
Therefore the contribution of the background dilaton concentrates
on the crosscap with the above gauge choice.

\section{Off-shell crosscap state}

In this section, we define the off-shell crosscap state and obtain
its explicit representation.
We apply the off-shell crosscap state to calculate 
the correlation functions and the partition function of 
the $RP^{2}$ worldsheet with quadratic background dilaton field.
We find that the behaviour of the off-shell crosscap state
signifies that a non-trivial dilaton condensation alters
the dimensions of O-planes.

\subsection{Off-shell crosscap conditions}
Let us consider an $RP^{2}$ worldsheet $\Sigma$ with the following
action:
\begin{eqnarray}
I&=&
\frac{1}{2\pi\alpha'}
\int_{\Sigma} d^{2} z \partial X^{\mu} \bar{\partial}X_{\mu}
+
\frac{1}{\pi}\int_{{\cal C}} d\sigma
\Phi(\sigma),
\label{ws-action}
\\
\Phi(\sigma)
&=&
a+\frac{1}{2\alpha'}\sum_{\mu=1}^{26}u_{\mu} (X^{\mu}(\sigma))^{2},
\label{boundary-term}
\end{eqnarray}
where the interaction $\Phi(\sigma)$ is inserted only on
crosscap ${\cal C}$.
Note that the worldsheet action is free in the ``bulk" region
$\{z=r e^{i\sigma} |0\le r < 1,0\le \sigma <2\pi\}$.
A closed string propagates freely in the ``bulk"  toward the crosscap, 
from the viewpoint of the closed-string channel.
In this sense, $X^{\mu}$ in the ``bulk"  can be expanded as
\begin{eqnarray}
X^{\mu}(z,\bar{z})
=
X^{\mu}_0-\frac{i\alpha'}{2}p^{\mu}\ln(z\bar{z})
+i\sqrt{\frac{\alpha'}{2}}
  \sum_{n\neq 0}
  \left(\alpha_n^{\mu}\frac{z^{-n}}{n}
     +\tilde{\alpha}_n^{\mu}\frac{\bar{z}^{-n}}{n} \right).
\end{eqnarray}

The existence of the crosscap, however, makes constraints on
the oscillation of the closed string in the neighborhood of
the crosscap (in the region $r \rightarrow 1$).
For example, the constraints when we have no interaction
on ${\cal C}$ are given in Ref. \cite{Callan-etal-2} as
\begin{eqnarray}
 \left\{
  X^{\mu}(z,\bar{z})-X^{\mu}(-1/\bar{z},-1/z)
 \right\}
\bigg|_{r \rightarrow 1}
&=&0,
\nonumber \\
 \left\{
  \dot{X^{\mu}}(z,\bar{z})+\dot{X^{\mu}}(-1/\bar{z},-1/z)
 \right\}
\bigg|_{r \rightarrow 1}
&=&0,
\label{well-known}
\end{eqnarray}
where
\begin{eqnarray}
\dot{X^{\mu}}(z,\bar{z})
&\equiv&
(w\partial_w + \bar{w}\bar{\partial}_{\bar{w}})X^{\mu}(w,\bar{w})
|_{w=z, \bar{w}=\bar{z}}
\:\:\:,
\nonumber \\
\dot{X^{\mu}}(-1/\bar{z},-1/z)
&\equiv&
(w\partial_w + \bar{w}\bar{\partial}_{\bar{w}})X^{\mu}(w,\bar{w})
|_{w=-1/\bar{z}, \bar{w}=-1/z}
\:\:\:.
\end{eqnarray}
Note that (\ref{well-known}) are equivalent to the following constraints:
\begin{eqnarray}
K_{0}(z,\bar{z})|_{r \rightarrow 1}
&=& 0,
\nonumber \\
\left\{
(z\partial_z + \bar{z}\bar{\partial}_{\bar{z}})
K_{0}(z,\bar{z})
\right\}|_{r \rightarrow 1}
&=& 0,
\label{K0}
\end{eqnarray}
where
\begin{eqnarray}
K_{0}(z,\bar{z})
&\equiv&
\dot{X}^{\mu}(z,\bar{z})+\dot{X}^{\mu}(-1/\bar{z},-1/z).
\end{eqnarray}
These conditions are rewritten in terms of closed-string modes
at $r \rightarrow 1$ as
\begin{eqnarray}
\alpha_n^{\mu}+(-1)^n \tilde{\alpha}_{-n}^{\mu}
&=&0,
\nonumber \\
p^{\mu}
&=&0.
\label{well-known-mode}
\end{eqnarray}
The conditions (\ref{well-known}), (\ref{K0}) or (\ref{well-known-mode})
are referred to as (on-shell) crosscap conditions.
 
The aim of this subsection is to extend the on-shell crosscap conditions
into the case $u_{\mu} \neq 0$.
We should find, in other words, the constraints on 
the closed-string modes in the neighborhood of ${\cal C}$
in the presence of interaction $\Phi$.
We call these constraints {\em off-shell crosscap conditions}.
We assert that the off-shell crosscap conditions can be
written as
\begin{eqnarray}
K(z,\bar{z})|_{r \rightarrow 1}
&=& 0,
\nonumber \\
\left\{
(z\partial_z + \bar{z}\bar{\partial}_{\bar{z}})
K(z,\bar{z})
\right\}|_{r \rightarrow 1}
&=& 0,
\label{gen-cross-cond}
\end{eqnarray}
where
\begin{eqnarray}
K(z,\bar{z})
&\equiv&
\dot{X}^{\mu}(z,\bar{z})+\dot{X}^{\mu}(-1/\bar{z},-1/z)
+
u_{\mu}\{X^{\mu}(z,\bar{z})+X^{\mu}(-1/\bar{z},-1/z)\}
\nonumber \\
&=&
\left.
\left\{
(w\partial_w + \bar{w}\bar{\partial}_{\bar{w}})
X^{\mu}(w,\bar{w})
+
u_{\mu}X^{\mu}(w,\bar{w})
\right\}
\right|_{w=z,\bar{w}=\bar{z}}
\nonumber \\
& &
+
\left.
\left\{
(w\partial_w + \bar{w}\bar{\partial}_{\bar{w}})
X^{\mu}(w,\bar{w})
+
u_{\mu}X^{\mu}(w,\bar{w})
\right\}
\right|_{w=-1/\bar{z},\bar{w}=-1/z}
.
\label{K}
\end{eqnarray}
The right-hand side of (\ref{K}) indicates the meaning of
the off-shell crosscap conditions;
(\ref{gen-cross-cond}) are the conditions so that
the $X^{\mu}$ in the neighborhood of ${\cal C}$, as well as 
its image by the involution, connects smoothly with the 
$X^{\mu}$ on ${\cal C}$
which obeys
\begin{eqnarray}
\left.
\left\{
(z\partial_z + \bar{z}\bar{\partial}_{\bar{z}})
X^{\mu}(z,\bar{z})+u_{\mu}X^{\mu}(z,\bar{z})
\right\}
\right|_{\cal C}
=0, 
\end{eqnarray}
given by varying the worldsheet action (\ref{ws-action}).\footnote{
Although $RP^2$ has no boundary,
$(z\partial_z + \bar{z}\bar{\partial}_{\bar{z}})X^{\mu}(z,\bar{z})$
which comes from the total derivative survives only on the
crosscap due to the involution.
}

The off-shell crosscap conditions (\ref{gen-cross-cond})
are rewritten in terms of closed-string modes as
\begin{eqnarray}
-\{
\alpha_n^{\mu}+(-1)^n \tilde{\alpha}_{-n}^{\mu}
\}
+
\frac{u_{\mu}}{n}
\{
\alpha_n^{\mu}-(-1)^n \tilde{\alpha}_{-n}^{\mu}
\}
&=&0,
\nonumber \\
-i\alpha' p^{\mu}+u_{\mu}X_0^{\mu}
&=&0,
\label{offshell-mode}
\end{eqnarray}
where we do not sum over $\mu$.
We can easily check that the conditions (\ref{offshell-mode}) gives
the on-shell crosscap conditions (\ref{well-known-mode})
in the limit $u^{\mu} \rightarrow 0$.
We also note that (\ref{offshell-mode}),
in the limit $u^{\mu} \rightarrow \infty$,
becomes equivalent to the T-dual of (\ref{well-known-mode}), 
\begin{eqnarray}
\alpha_n^{\mu}-(-1)^n \tilde{\alpha}_{-n}^{\mu}
&=&0,
\nonumber \\
X_0^{\mu}
&=&0,
\end{eqnarray}
which are rewritten as the T-dual of (\ref{well-known}):
\begin{eqnarray}
 \left\{
  X^{\mu}(z,\bar{z})+X^{\mu}(-1/\bar{z},-1/z)
 \right\}
\bigg|_{r \rightarrow 1}
&=&0,
\nonumber \\
 \left\{
  \dot{X^{\mu}}(z,\bar{z})-\dot{X^{\mu}}(-1/\bar{z},-1/z)
 \right\}
\bigg|_{r \rightarrow 1}
&=&0.
\label{well-known-T}
\end{eqnarray}
Note that the conditions (\ref{gen-cross-cond}) in the limit 
$u^{\mu} \rightarrow \infty$ is (\ref{well-known-T}) itself.
Therefore the off-shell crosscap conditions 
(\ref{gen-cross-cond}) or (\ref{offshell-mode})
naturally interpolate between on-shell crosscap conditions 
and their T-duals.
The coupling constant $u^{\mu}$, which is a parameter of the 
configuration of the background dilaton field, controls 
the dimension of the corresponding O-plane.

\subsection{Off-shell crosscap state and  partition function}
We define off-shell crosscap state
$\langle C(\mbox{\boldmath$u$})|$ 
using the off-shell crosscap conditions as
\begin{eqnarray}
\langle C(\mbox{\boldmath$u$})|
\left\{
-\{
\alpha_n^{\mu}+(-1)^n \tilde{\alpha}_{-n}^{\mu}
\}
+
\frac{u_{\mu}}{n}
\{
\alpha_n^{\mu}-(-1)^n \tilde{\alpha}_{-n}^{\mu}
\}
\right\}
&=&0,
\nonumber \\
\langle C(\mbox{\boldmath$u$})|
\left\{
-i\alpha' p^{\mu}+u_{\mu}X_0^{\mu}
\right\}
&=&0.
\label{boundary-cond}
\end{eqnarray}
The explicit form of $\langle C(\mbox{\boldmath$u$})|$ is
given as
\begin{eqnarray}
\langle C(\mbox{\boldmath$u$})| &=& \langle 0|{\cal C}(\mbox{\boldmath$u$}),
\nonumber \\
{\cal C}(\mbox{\boldmath$u$})&\equiv&
\exp \left(-\frac{1}{2}X_0^{\mu} A_{\mu \nu} X_0^{\nu} \right)
\exp \left(\sum_{m=1}^{\infty} \tilde{\alpha}_m^{\mu}
    C_{\mu \nu}^{(m)} \alpha_m^{\nu} \right),
\label{boundary-state}
\end{eqnarray}
where
\begin{eqnarray}
A_{\mu \nu}
\equiv A(u_{\mu})\delta_{\mu \nu}
=
\frac{1}{\alpha'}u_{\mu} \delta_{\mu \nu},
\:\:\:\:
C_{\mu \nu}^{(m)}
\equiv
C^{(m)}(u_{\mu})\delta_{\mu \nu}
=
-\frac{(-1)^m}{m}\frac{m-u_{\mu}}{m+u_{\mu}}\delta_{\mu \nu}.
\end{eqnarray}
We can easily check that this off-shell crosscap state becomes (the
T-dual of) the usual on-shell crosscap state if we take the limit
$u^{\mu} \rightarrow 0$ ($u^{\mu} \rightarrow \infty$).
Therefore the off-shell crosscap state naturally interpolates between
the crosscap state for a higher-dimensional O-plane and that for a
lower-dimensional O-plane.

Next, we show that the off-shell crosscap state is a useful tool to
evaluate the quantities on the $RP^{2}$ worldsheet.
For example, we can calculate the Green's function and the partition
function on the $RP^{2}$ worldsheet in the presence of interaction
$\Phi(\sigma)$ on the crosscap.
Let us consider one-dimensional target space and omit the superscript
$\mu$ of $X$ and $u$ for simplicity.
The Green's function for this case is given as
\begin{eqnarray}
G(z,w)&=&
\frac{\langle C(u)| X(z,\bar{z}) X(w,\bar{w})|0\rangle}{\langle
C(u)|0\rangle}
\nonumber \\
&=&
-\frac{\alpha'}{2}\ln|z-w|^{2} -\frac{\alpha'}{2}\ln|1+z\bar{w}|^{2}
\nonumber \\
& &+\frac{\alpha'}{u}
-\alpha' u \sum_{k=1}^{\infty}\frac{1}{k(k+u)}
\left[(-z\bar{w})^k+(-\bar{z}w)^k\right].
\end{eqnarray}
In the case $z=e^{i\sigma}$ and $w=e^{i\sigma'}$, we obtain
\begin{eqnarray}
G(e^{i\sigma},e^{i\sigma'})
&=&
-\frac{\alpha'}{2}\ln|1-e^{i(\sigma-\sigma')}|^{2}
-\frac{\alpha'}{2}\ln|1+e^{i(\sigma-\sigma')}|^{2}
\nonumber \\
& &+\frac{\alpha'}{u}
-\alpha' u \sum_{k=1}^{\infty}\frac{(-1)^{k}}{k(k+u)}
\left[e^{ik(\sigma-\sigma')}+e^{-ik(\sigma-\sigma')}\right].
\end{eqnarray}
We define composite operator $X^{2}(\sigma)$ as 
shown in Ref. \cite{wittenbsft-2}:
\begin{eqnarray}
X^{2}(\sigma)
&\equiv&
\lim_{\epsilon \rightarrow 0}
\left(
X(\sigma)X(\sigma+\epsilon)-f(\epsilon)
\right),
\nonumber \\
f(\epsilon)
&=&-\frac{\alpha'}{2}\ln|1-e^{i\epsilon}|^{2}
+ (\mbox{const.}) .
\label{def-X2}
\end{eqnarray}
We write the constant in (\ref{def-X2}) as $\alpha' \ln q$
by using a positive constant $q$.
The value for $q$ is ambiguous at this stage and depends on the
renormalization scheme.
We will determine the value for $q$ later.
We can then obtain
\begin{eqnarray}
\langle X^{2}(\sigma) \rangle
&=&
-\frac{\alpha'}{2}\ln2^{2} +\frac{\alpha'}{u}
-2\alpha' u \sum_{k=1}^{\infty}\frac{(-1)^{k}}{k(k+u)}
-\alpha' \ln q
\nonumber \\
&=&
-\alpha' \ln(2q) -\frac{\alpha'}{u}
-2\alpha' \left[ \Psi\left(\frac{u}{2}\right)-\Psi(u) \right],
\label{X2}
\end{eqnarray}
where
\begin{eqnarray}
\Psi(u) \equiv \frac{\frac{d}{du}\Gamma(u)}{\Gamma(u)}
\end{eqnarray}
is a polygamma function for which we have used the relationship
\begin{eqnarray}
u \sum_{k=1}^{\infty}\frac{(-1)^{k}}{k(k+u)}
=
\frac{1}{u}+\Psi\left(\frac{u}{2}\right)-\Psi(u).
\end{eqnarray}

We can next calculate the partition function of the $RP^{2}$
worldsheet by using the following relationship:
\begin{eqnarray}
\frac{d}{du}\ln Z(u)
&=&
-\frac{1}{2\pi\alpha'}
\int_{0}^{\pi} d\sigma
\langle X^{2}(\sigma) \rangle
=
-\frac{1}{2\alpha'}\langle X^{2}(\sigma) \rangle
\nonumber \\
&=&
-\frac{1}{2\alpha'}
\left\{
  -\alpha' \ln(2q) -\frac{\alpha'}{u}
  -2\alpha' \left[ \Psi\left(\frac{u}{2}\right)-\Psi(u) \right]
\right\}
\nonumber \\
&=&
\frac{1}{2}
\left\{
\ln(2q)+\frac{d}{du}\ln u
  +2\left[
2\frac{d}{du}\ln\Gamma\left(\frac{u}{2}\right)-\frac{d}{du}\ln\Gamma(u)
\right]
\right\}
\nonumber \\
&=&
\frac{d}{du}
\left(
\ln
  \frac{{\sqrt{2q}}^u \sqrt{u}
{\Gamma\left(\frac{u}{2}\right)}^{2}}{\Gamma(u)}
  + \:\:\: \mbox{const.}
\right).
\end{eqnarray}
We then obtain
\begin{eqnarray}
Z(u)
=
  \frac{{\sqrt{2q}}^u \sqrt{u} {\Gamma(\frac{u}{2})}^{2}}{\Gamma(u)}
,
\label{Z(u)}
\end{eqnarray}
up to the overall normalization factor.
In general, the partition function for 26-dimensional target space with
the interaction (\ref{boundary-term}) on the crosscap can be written as
\begin{eqnarray}
Z(a,\mbox{\boldmath$u$})
&\equiv&
e^{-a} Z(\mbox{\boldmath$u$})
\equiv
e^{-a} \prod_{\mu=1}^{26} Z(u_{\mu})
\nonumber \\
&=&
e^{-a} \prod_{\mu=1}^{26}
{\left(
  \frac{{\sqrt{2q}}^{u_{\mu}}
  \sqrt{u_{\mu}} {\Gamma(\frac{u_{\mu}}{2})}^{2}}{\Gamma(u_{\mu})}
\right)},
\label{general-Z(u)}
\end{eqnarray}
up to the overall normalization factor.
We note that the partition function (\ref{general-Z(u)}) on the
$RP^{2}$ has an identical representation with the partition function
on the supersymmetric disc worldsheet which have been considered
in the analysis of open-string tachyon condensation in $D\bar{D}$
systems \cite{0010108,0012198,0012210}.
We present some comments on the relationship between the $RP^{2}$ 
worldsheet and the supersymmetric disc worldsheet in Appendix.

\section{Derivation of the descent relation among O-plane tensions}

In this section, we clarify the physical meaning of the partition 
function calculated in the previous section.
We show that the condensation of the quadratic dilaton field
reproduces the correct descent relation among the O-plane tensions.

\subsection{Sigma model approach and $RP^{2}$ worldsheet}

Let us consider our work from the viewpoint of the sigma model approach.
The basic idea of the sigma model approach is that the spacetime action
for string fields is essentially the renormalized partition 
function of the worldsheet with corresponding background string fields.
In this sense, the spacetime action $S$ for string field $\lambda_i$
may be given as
\begin{eqnarray}
S(\lambda_i)
&\cong&
\sum_{\chi}\int_{\Sigma_{\chi}}[d g_{ab}][dX^{\mu}]
e^{-I_{\chi}(g_{ab},X^{\mu};\lambda_i)}
\nonumber \\
&=&
Z_{sphere}(\lambda_i)
+\{Z_{disc}(\lambda_i)+Z_{RP^{2}}(\lambda_i)\}+\cdots
,
\label{sigma}
\end{eqnarray}
where $I_{\chi}$ is the action on the worldsheet $\Sigma_{\chi}$ 
of the Euler number $\chi$.
Leading term $Z_{sphere}(\lambda_i)$ is the renormalized partition
function on the sphere. This term is of order $g_o^{-2}$ 
where $g_o$ is the open-string coupling constant.
Renormalized partition functions
$Z_{disc}(\lambda_i)$ on the disc and $Z_{RP^{2}}(\lambda_i)$  
on the $RP^{2}$
are the loop correction terms of order $g_o^{-1}$.
In principle, $Z_{disc}(\lambda_i)$ is proportional to the tension of
the corresponding D-brane and $Z_{RP^{2}}(\lambda_i)$ is proportional
to the tension of the corresponding O-plane.

It is known, however, that the right-hand side of (\ref{sigma}) 
does not give the correct spacetime action;
we need modification of $Z_{sphere}(\lambda_i)$ and 
$Z_{disc}(\lambda_i)$
\footnote{
We need not modify $Z_{disc}(\lambda_i)$ in superstring theory since
the volume of the super-M\"{o}bius group of the super-disc is finite
\cite{Mobius-1}.
}
in order to obtain the correct off-shell spacetime action.
These modifications are closely related to the infinite 
M\"{o}bius volume of the worldsheets;
we need to subtract the divergence from the M\"{o}bius infinity
\cite{Mobius-1,Mobius-2,Mobius-3}.
(See also \cite{tseytlin11/2000,0010108}.)

The situation is different for $Z_{RP^{2}}(\lambda_i)$
(and for the partition functions of the worldsheets of $\chi \le 0$).
We should note that the M\"{o}bius group of $RP^{2}$ is $SO(3)$ whose
volume is finite, and we have no M\"{o}bius infinity from the $RP^{2}$ 
worldsheet.
Therefore it is natural to assume that partition function
$Z_{RP^{2}}(\lambda_i)$ itself is the exact loop correction term
from the $RP^{2}$ graph.
We have calculated the partition function (\ref{general-Z(u)})
on the $RP^{2}$ worldsheet in the presence of quadratic background
dilaton field $\Phi=a+(2\alpha')^{-1}\sum_{\mu}u_{\mu} X_{\mu}^{2}$ 
on the crosscap.
We have fixed the worldsheet metric so that the ``bulk"  part of the 
fundamental region of the $RP^2$ becomes flat and the contribution 
of the dilatons concentrates on the crosscap.
Therefore we have to calculate the contribution of the ghost field 
and the anti-ghost field on the $RP^{2}$ worldsheet in order to
obtain the correct overall normalization of (\ref{general-Z(u)}).
We write this overall factor $A$ and then we have the following 
relationship:
\begin{eqnarray}
Z_{RP^{2}}(\Phi)
&=&A
e^{-a} \prod_{\mu=1}^{26}
{\left(
  \frac{{\sqrt{2q}}^{u_{\mu}}
  \sqrt{u_{\mu}} {\Gamma(\frac{u_{\mu}}{2})}^{2}}{\Gamma(u_{\mu})}
\right)}
\nonumber \\
&\equiv&
Z_{RP^{2}}(a,\mbox{\boldmath$u$}).
\end{eqnarray}
The sign of A should be minus.
Note that factor $e^{-a}$ is equal to $g_{o}^{-1}$ when
the dilaton is constant.

\subsection{Asymptotic behaviour of $Z(u)$}

Let us rewrite $Z(u)$ given in (\ref{Z(u)}) as
\begin{eqnarray}
Z(u)
&=&
\frac{{\sqrt{2q}}^u \sqrt{u} {\Gamma(\frac{u}{2})}^{2}}{\Gamma(u)}
=
\frac{4}{\sqrt{u}} \left(\frac{q}{2}\right)^{\frac{u}{2}}
F\left(\frac{u}{2}\right),
\end{eqnarray}
where we have defined function $F(x)$ as
\begin{eqnarray}
F(x)
\equiv
\frac{4^x x{\Gamma(x)}^{2}}{2\Gamma(2x)}.
\label{F(x)}
\end{eqnarray}
$F(x)$ behaves as follows:
\begin{eqnarray}
F(x)
&\sim&
1+(2\ln 2)x + O(x^{2})
\:\:\:\: (x \rightarrow 0),\\
F(x)
&\sim&
\sqrt{\pi x} + O(x^{-\frac{1}{2}})
\:\:\:\: (x \rightarrow \infty).
\end{eqnarray}
$Z(u)$ around $u=0$ is then
\begin{eqnarray}
Z(u)
&\sim& 4
\left( \frac{1}{\sqrt{u}} \right)
+O(\sqrt{u}).
\label{Z(u)-lim-0}
\end{eqnarray}
Thus, $Z(u)$ diverges when $u$ approaches $0$. 
This is an IR divergence which corresponds to the volume of the spacetime.

On the other hand
$Z(u)$ around $u=\infty$ is
\begin{eqnarray}
Z(u)
&\sim&
\left(\frac{q}{2}\right)^{\frac{u}{2}}
\left\{4 \sqrt{\frac{\pi}{2}} + O\left(\frac{1}{\sqrt{u}}\right) \right\}
.
\end{eqnarray}
We can obtain a finite and non-zero value of $Z(u)$ in the limit
$u\rightarrow \infty$ if and only if $q=2$.
Therefore we assign the value $2$ to $q$.
In other words, we have chosen the renormalization scheme in 
(\ref{def-X2}) so that we can obtain a finite and non-zero value 
of $Z(u)$ in the limit $u\rightarrow \infty$.
$Z(u)$ is then
\begin{eqnarray}
Z(u)
=
\frac{{\sqrt{4}}^u \sqrt{u} {\Gamma(\frac{u}{2})}^{2}}{\Gamma(u)}
=
\frac{4}{\sqrt{u}}
F\left(\frac{u}{2}\right),
\label{Z(u)-in-F}
\end{eqnarray}
up to the overall normalization factor and
\begin{eqnarray}
Z_{RP^{2}}(\mbox{a,\boldmath$u$})
=
Ae^{-a}
\prod_{\mu=1}^{26}
\left(
 \frac{4}{\sqrt{u_{\mu}}}
F\left(\frac{u_{\mu}}{2}\right)
\right).
\label{Z(bold-u)}
\end{eqnarray}

\subsection{Ratio of the O-plane tensions }

Let us define quantity $S_p$ as follows:
\begin{eqnarray}
Z_{RP^{2}}(\mbox{a,\boldmath$u$})
\rightarrow
S_p,
\end{eqnarray}
where the limit is taken as
\begin{eqnarray}
u^{1},\cdots, u^{p+1} &\to& 0, \nonumber \\
u^{p+2},\cdots, u^{26} &\to& \infty.
\end{eqnarray}
We do not touch the parameter $a$ in this subsection.
According to the argument in the previous section,
$S_p$ is equal to $V_p \times T_p$ where
$V_p$ and $T_p$ are the volume and tension of an O$p$-plane.
Here, the dimension of the O-plane is $p+1$ and is defined as 
the number of parameters $u^{\mu}$ which are taken to zero.
We can thus write
\begin{eqnarray}
S_{25}&=&
Z_{RP^{2}}(u_{25} \to 0,u_{i} \to 0)=\int dx^{25} V_{24} T_{25},
\nonumber \\
S_{24}&=&
Z_{RP^{2}}(u_{25} \to \infty,u_{i} \to 0)=V_{24} T_{24},
\end{eqnarray}
where $i\neq 25$.
Then
\begin{eqnarray}
\frac{S_{25}}{S_{24}}
=
\frac{Z(u_{25} \to 0)}{Z(u_{25} \to \infty)}
=
\frac{\int dx^{25} T_{25}}{T_{24}},
\end{eqnarray}
where $Z(u)$ is given in (\ref{Z(u)-in-F}).
We can rewrite $Z(u_{25})$ as
\begin{eqnarray}
Z(u_{25})&=&\frac{4}{\sqrt{u_{25}}} F\left(\frac{u_{25}}{2}\right)
\nonumber \\
&=&\int dx^{25} e^{-\frac{1}{2\alpha'}u_{25}(x^{25})^{2}}
  \frac{1}{\sqrt{2\alpha'\pi}}
  4F\left(\frac{u_{25}}{2}\right),
\end{eqnarray}
where $x^{25}$ is the zero mode of $X^{25}$.
In the second line, we have explicitly rewritten the integral 
of the zero-mode part of
$Z(u_{25})$ \cite{0010108,Nakamura}.
(This corresponds to $\sim \frac{1}{\sqrt{u_{25}}}$.)
We can therefore obtain
\begin{eqnarray}
\lim_{u_{25} \rightarrow 0}Z(u_{25})
=
\lim_{u_{25} \rightarrow 0}
\int dx^{25} e^{-\frac{1}{2\alpha'}u_{25}(x^{25})^{2}}
  \frac{1}{\sqrt{2\alpha'\pi}}
  4F\left(\frac{u_{25}}{2}\right)
=
\int dx^{25} \frac{4}{\sqrt{2\alpha'\pi}} \cdot 1.
\end{eqnarray}
We can also obtain
\begin{eqnarray}
\lim_{u_{25} \rightarrow \infty}Z(u_{25})
&=&
\lim_{u_{25} \rightarrow \infty}
\frac{4}{\sqrt{u_{25}}} F\left(\frac{u_{25}}{2}\right)
=
\lim_{u_{25} \rightarrow \infty}
\frac{4}{\sqrt{u_{25}}}
\left( \sqrt{\pi \frac{u_{25}}{2}}+O\left(\frac{1}{\sqrt{u_{25}}}\right)
\right)
\nonumber \\
&=&
4\sqrt{\frac{\pi}{2}}.
\end{eqnarray}
Therefore
\begin{eqnarray}
\frac{T_{24}}{T_{25}}=\frac{\sqrt{2\alpha'\pi}}{4} 4\sqrt{\frac{\pi}{2}}
=\pi \sqrt{\alpha'}.
\end{eqnarray}
This is precisely the ratio of the tension of an O$24$-plane and
that of an O$25$-plane.
In general, we can show in a similar manner that
\begin{eqnarray}
\frac{T_{p}}{T_{q}}=\left( \pi \sqrt{\alpha'} \right)^{q-p}.
\end{eqnarray}

\section{Conclusion}

We have considered the relationship between the configuration of the
background dilaton field and the dimensions of the O-planes.
We showed that the contribution of the dilatons on the $RP^{2}$
worldsheet localizes on the crosscap if we choose the gauge in which
the worldsheet curvature in the ``bulk" vanishes.
This feature enables us to treat dilatons quite easily.
We have proposed the off-shell crosscap state which naturally
interpolates between the usual crosscap states and their T-duals.
The behaviour of the off-shell crosscap state signifies that
the non-trivial dilaton condensation alters the dimensions of
O-planes.
We obtained the correlation functions and partition function
on the $RP^{2}$ worldsheet in the presence of the quadratic dilaton
field on the crosscap.
We showed that the non-trivial dilaton condensation reproduces
the correct descent relation among O-plane tensions, 
by taking the on-shell limits of the partition function of the $RP^{2}$.
We found the correspondence between the $RP^{2}$ worldsheet 
and the supersymmetric disc which is presented in Appendix.

We would like to make some comments.
We have studied the effects of the non-trivial 
dilaton condensation on O-planes.
In order to describe the full dynamics of dilatons,
we would need open-closed string field theory \footnote{
An unoriented open-closed string field theory has been proposed in Ref.
\cite{closed-SFT}.
}. 
Non-perturbative analysis would be necessary too.
However the present work can be a step to understand the relationship
between the condensation of string fields and the transmutations
of O-planes.
An extension of the present work into the supersymmetric case is
interesting, and studying the relationship between our work and the
properties of O-planes
described by F-theory \cite{Sen-1,Dab-1} is one further direction 
to pursue.
Studying the relationship between the configuration of the dilatons and
the O-planes in Type I' theory \cite{Pol-witten,Arakane-etal} 
is also tempting.
It has been shown that non-trivial configurations of dilatons
constrain the spacetime positions of D-branes \cite{Nakamura-1}.
Effects of the non-trivial dilaton condensation on D-branes are
also interesting subjects to study.
Recent studies have shown that the condensation of the closed-string
tachyons in the twisted sector can alter the topology of the orbifolded
spacetime 
\cite{Don'tPanic,Vafa-Mirror,LocalizedTachyons,Dab-Vafa}.
We expect that the present work can also be a step towards understanding
the relationship between the topology of the spacetime and the configuration
of the string fields, since the dimensions of O-planes are closely
related to the topology of the orientifolded spacetime.

\vspace{1cm}
\noindent
{\large\bf Acknowledgment}

We would like to thank H. Kawai, T. Kugo, K. Murakami, T. Nakatsu, 
A. Tsuchiya and T. Yokono for fruitful discussions.
We are grateful to Santo seminar for providing an opportunity for
collaboration.
S.N. wishes to thank M. Laidlaw, M. Rozali, and  G. W. Semenoff  for
discussions and their hospitality during his stay at University of
British Columbia.

\newpage
\appendix

\setcounter{equation}{0}

\section{A}
\noindent
{\large\bf Correspondence between the $RP^{2}$ worldsheet 
and the supersymmetric disc worldsheet}\\

We comment in this appendix on the relationship between the $RP^{2}$
worldsheet we have considered and the supersymmetric disc (super-disc)
worldsheet which appears in the analysis of $D\bar{D}$ systems.
Let us consider a unit disc
${\cal M}$ ($\{z=re^{i\sigma}|0\le r \le 1, 0\le \sigma <2\pi\}$) 
with the following worldsheet action:
\begin{eqnarray}
I_{super-disc}&=&
\frac{1}{4\pi}\int_{{\cal M}} d^{2} z
\left\{
\frac{2}{\alpha'}\partial X^{\mu} \bar{\partial} X_{\mu}
+\psi^{\mu}\bar{\partial}\psi_{\mu}
+\tilde{\psi}^{\mu}\partial\tilde{\psi}_{\mu}
\right\}  \nonumber
\\
&&+
\frac{1}{4\pi}\int_{\partial {\cal M}} d\sigma
y^{\mu}
\left\{
\frac{2}{\alpha'}X_{\mu}^{2}
+(\psi^{\mu}\partial_{\mu}X^{\rho})\frac{1}{\partial_{\sigma}}
 (\psi^{\nu}\partial_{\nu}X_{\rho})
\right\}.
\label{super-ws-action}
\end{eqnarray}
This action has been considered in the context of open-string tachyon
condensation in $D\bar{D}$ systems \cite{0010108,0012198,0012210}.
The partition function of the super-disc is obtained in
Ref. \cite{0010108} as
\begin{eqnarray}
Z(\mbox{\boldmath$y$})
\propto
\prod_{\mu=1}^{26}\sqrt{\frac{1}{y^{\mu}}}F(y^{\mu}),
\label{Z(y)}
\end{eqnarray}
where function $F(x)$ has been defined in (\ref{F(x)}).

Therefore, the partition function (\ref{Z(bold-u)}) of the $RP^{2}$
worldsheet has an identical form to the partition function (\ref{Z(y)})
of the super-disc.
Comparing (\ref{Z(bold-u)}) and (\ref{Z(y)}), 
we find the correspondence
\begin{eqnarray}
\frac{u^{\mu}}{2}
\leftrightarrow
y^{\mu}.
\end{eqnarray}
At the level of integrated operators, we find the correspondence
\begin{eqnarray}
\label{lrarrow}
O_{RP^{2}}  \leftrightarrow  O_{disc} ,
\end{eqnarray}
where
\begin{eqnarray}
O_{RP^{2}}
\equiv  \int_{\cal C} d\sigma
  \frac{2}{\alpha'}  X_{\mu}^{2}(\sigma)
\end{eqnarray}
is an operator on the $RP^{2}$ and
\begin{eqnarray}
O_{disc}(\sigma)
\equiv  \int_{\partial {\cal M}} d\sigma
\left\{
\frac{2}{\alpha'} X_{\mu}^{2}(\sigma)
+(\psi^{\mu}\partial_{\mu}X^{\rho})\frac{1}{\partial_{\sigma}}
 (\psi^{\nu}\partial_{\nu}X_{\rho})(\sigma)
\right\}
\end{eqnarray}
is an operator on the super-disc.
We should also note that the descent relation among D-brane tensions 
in the $D\bar{D}$ system can be obtained in the same manner
as that we have shown in Sec. 4.
The correspondence (\ref{lrarrow}) can be explained from
the calculation of $\langle X_{\mu}^{2} \rangle$ for the $RP^{2}$,
in which we have an extra minus sign in the contributions 
from odd modes.\footnote{
See the third term of the first line on 
the right-hand side of (\ref{X2}).}
These contributions correspond to those of the fermions on 
the super-disc, while the even modes
behave like the bosonic part on the super-disc.


\end{document}